\PassOptionsToPackage{table}{xcolor}
\documentclass[twocolumn,prd,preprintnumbers,numbers,sort&compress,nofootinbib,showpacs,colorlinks,citecolor=blue,amsmath,amssymb,aps,superscriptaddress]{revtex4-2}

\usepackage{graphicx}% Include figure files
\usepackage{dcolumn}% Align table columns on decimal point
\usepackage{bm}% bold math
\usepackage{xcolor}
\usepackage{lipsum} 
\usepackage{float} 
\usepackage[utf8]{inputenc}

\newcommand{\beq}{\begin{equation}}
\newcommand{\eeq}{\end{equation}}
\newcommand{\exclude}[1]{}

\usepackage{amsmath}
\usepackage[caption=false]{subfig}

\newcommand{\be}{\begin{eqnarray}}
\newcommand{\ee}{\end{eqnarray}}

\def\+{\dagger}

\def\ra{\rangle}
\def\<{\langle}
\def\>{\rangle}

\newcommand{\Lqcd}{\Lambda_{\mathrm{QCD}}}

\newcommand{\Tr}{\mathrm{Tr}}

\newcommand{\ubc}{Department of Physics \& Astronomy, University of British Columbia, 6224 Agricultural Road, Vancouver, BC V6T 1Z1, Canada}

\usepackage{hyperref}

\begin{document}

\title{DESI results and  Dark Energy from QCD topological sectors}

\author{Ludovic van Waerbeke}%
 \email{waerbeke@phas.ubc.ca}
\affiliation{\ubc}

%\author[orcid=0000-0002-2637-8728]{Ludovic Van Waerbeke}

\author{Ariel Zhitnitsky} 
%\altaffiliation{Las Campanas Observatory}
\email{arz@phas.ubc.ca}
\affiliation{\ubc}

%\collaboration{all}{The Terra Mater collaboration}

\begin{abstract}
We present a physically motivated model of dark energy (DE) rooted in the topological structure of the Quantum Chromodynamics (QCD) vacuum. In this framework, DE arises from the difference between the QCD vacuum energy in an expanding Universe and in Minkowski spacetime, with the relevant contribution generated by tunnelling between distinct topological sectors. The QCD scale, $\Lambda_{\rm QCD}\sim100,{\rm MeV}$, naturally produces a vacuum-energy density of the same order of magnitude as the observed one, without introducing new fundamental fields, couplings, or fine-tuning. We show that the same mechanism can naturally accommodate an evolving dark-energy component. The resulting effective equation-of-state parameter $\omega_{\rm DE}$ can evolve with redshift, lie on either side of $-1$, and cross the $\omega_{\rm DE}=-1$ boundary, while asymptotically approaching $-1$ in the future. The framework offers a common physical origin for both the small amplitude of dark energy and its possible time evolution, the latter being suggested by recent DESI observations. Here, we develop the physical and cosmological aspects of the framework; its confrontation with cosmological data is presented in a companion paper.

\end{abstract}

%\keywords{\uat{Cosmology}{343} --- \uat{High Energy astrophysics}{739}}

\maketitle
\section{Introduction}\label{introduction}

This paper addresses two central aspects of the dark-energy problem: the origin of its small amplitude and the possibility, suggested by recent observations, that its equation of state evolves with time.

Since the discovery of the accelerated expansion of the Universe, attributed to dark-energy (DE) \citep{1999ApJ...517..565P,1998AJ....116.1009R}, a wide variety of models have been proposed to explain its physical origin. To this day, none has resolved the ``cosmological constant problem'': the calculated vacuum self-energy of quantum fields is is $\sim 120$ orders of magnitude larger than the observed DE density.

The recent DESI results \citep{2025arXiv250314738D} provide some of the most precise constraints to date on the DE equation-of-state parameter, $\omega_{\rm DE}$. Far from resolving the cosmological constant problem, however, these results could add a new layer of complexity. When combined with other cosmological probes\footnote{Cosmic Microwave Background (CMB) and Supernovae (SNIa) data.}, DESI favors an evolution in which $\omega_{\rm DE}>-1$ at the present epoch, while crossing below $-1$ at higher redshift. Such an evolution, if confirmed, would pose a severe theoretical challenge, as any successful model DE would have to account not only for its small amplitude, but also for its apparent time dependence and crossing of the $\omega_{\rm DE}=-1$ boundary. The regime $\omega_{\rm DE}<-1$, known as the ``phantom regime'', is theoretically contrived, since its realization within conventional quantum field theory (QFT) generally requires degrees of freedom with pathological properties, potentially leading to violations of unitarity or vacuum stability. An evolution across the $\omega_{\rm DE}=-1$ boundary (the so-called phantom crossing), poses an additional difficulty \citep{Cai:2009zp,2026NSRev..13ag115C}: such a crossing cannot be consistently realized by a single minimally coupled canonical scalar field and, in more general QFT constructions, is often accompanied by instabilities or other pathologies, such as ghost degrees of freedom or gradient instabilities associated with an imaginary speed of sound $c_s^2<0$. These considerations raise fundamental objections to describing an evolving DE component that crosses $\omega_{\rm DE}=-1$ in terms of a conventional dynamical canonical field $\Phi$ governed by a potential $V(\Phi)$. On the other hand, an evolving DE component has been argued for theoretically \citep{2025arXiv250317659B}.

%From this perspective, the DE problem has two distinct aspects: first, its observed amplitude is extraordinarily small compared with the naive theoretical expectation for vacuum energy; second, accounting for its possible time evolution in terms of conventional propagating fields introduces additional theoretical difficulties. Together, these issues make a description of dark energy solely in terms of local propagating degrees of freedom particularly challenging.

\bigskip

These problems can be mitigated if one accepts the Zel'dovich proposal formulated in 1967 \citep{Zeldovich:1967gd}, long before the DE was discovered. The basic idea is that the self-energy contribution of virtual particles might somehow cancel, while their mutual gravitational interaction could leave a much smaller residual vacuum energy, $\rho_{\rm vac}$. He estimated this gravitational contribution to be of order
\be
\rho_{\rm vac}\sim \frac{Gm_p^6c^4}{\hbar^4},
\label{eq:Zeldo}
\ee
where $m_p$ is the proton mass, $c$ is the speed of light, and $\hbar$ is the reduced Planck constant. In this picture, the relevant DE contribution originates from the coupling between the quantum vacuum and the gravitational background. The physical significance and historical context of this proposal are discussed in detail in the review \cite{2008GReGr..40.1557S}. Zel'dovich's argument substantially reduces the severity of the vacuum-energy problem, replacing the discrepancy $\sim 120$ orders of magnitude with one of only a few orders of magnitude. Most importantly, it opens the possibility that the observed cosmological constant originates from an interplay between gravity and particle physics, rather than from a purely quantum-field-theoretic vacuum contribution.

Building on a modern formulation of Zel'dovich's proposal, \cite{Zhitnitsky:2013pna,Zhitnitsky:2015dia,Barvinsky:2017lfl} suggested defining the gravitationally relevant vacuum energy as the difference between $\rho_{\rm ST}$, the vacuum energy in a generic curved and/or time-dependent spacetime, and $\rho_{\mathrm{Mink}}$, the corresponding vacuum energy in Minkowski spacetime:
\be
\rho_{\rm vac}\equiv \rho_{\rm ST}-\rho_{\mathrm{Mink}}.
\label{eq:DEdef}
\ee
Minkowski spacetime provides a natural reference background in which both curvature and cosmic expansion are absent. In this prescription, the large spacetime-independent contribution to the vacuum energy is removed by subtraction, while the remaining term measures the response of the quantum vacuum to the non-trivial gravitational background. In this sense, Eq.~(\ref{eq:DEdef}) provides a concrete realization of Zel'dovich's original proposal.

In the present work, we will develop a framework to promote this theoretical idea into a concrete and testable cosmological model applicable throughout the evolution of the Universe. We will focus exclusively on the construction of the formalism, while its confrontation with cosmological observations is presented in a companion paper \cite{2026arXiv260620036L}. We will show that this framework addresses both central aspects of the DE problem: its observed amplitude and its possible time evolution.

The paper is organized as follows. In Section~\ref{sec:topology}, we introduce the central idea. Sections~\ref{intuition} presents an example of non-local effects in gauge field theories. Section~\ref{sec:lessintuition} presents a mathematically exact calculation of the induced dark-energy contribution in a QCD-inspired toy model. Section~\ref{holonomy} summarizes the core conceptual features of the framework. Section~\ref{sect:EoS} describes the de Sitter limit. In Section~\ref{DESI}, we derive the solution to the Friedmann equation within our framework and analyze its behavior both in the asymptotic limit, $t\rightarrow\infty$, and over the redshift range $z\geq 0$. We then discuss several possible cosmological implications of the model and the ways in which it can be tested, before concluding in Section~\ref{conclusion}.

\section{Topological Origin of Gravitating Vacuum Energy}
\label{sec:topology}

The purpose of this section is to outline how the vacuum energy of the matter sector in a non-Minkowski spacetime may contribute to the cosmological dark-energy background.

The basic ideas were developed in \cite{Zhitnitsky:2013pna,Zhitnitsky:2015dia,Barvinsky:2017lfl}, where the matter sector is described by Quantum Chromodynamics (QCD) and the background cosmology by a curved hyperbolic spacetime, where computations can be carried out. A remarkable result of these studies is that the QCD vacuum energy density in curved spacetime, $\rho_{\rm ST}$, contains two distinct contributions: a local component associated with propagating degrees of freedom, such as particles and fields, and a genuinely non-local component associated with the non-trivial topological structure of the QCD vacuum. After subtracting the corresponding Minkowski vacuum energy density, the local contribution cancels exactly, leaving only the non-local component. This surviving term is an infrared\footnote{Here “infrared” refers to the very-long-distance, or equivalently very-small-momentum, sector of the theory.} contribution to the background energy density and may be interpreted as a global vacuum-energy contribution of the theory. Extrapolating this result to a de Sitter Universe, \cite{Zhitnitsky:2013pna,Zhitnitsky:2015dia,Barvinsky:2017lfl} showed that the resulting DE density naturally falls within the same order of magnitude as today's observed value.

A common objection to a cosmological role played by QCD is that its characteristic interaction scale is microscopic and hence cannot play any cosmological role. This is true for effects mediated by propagating degrees of freedom: because QCD has a mass gap\footnote{i.e. the vacuum is separated from the lightest physical excitation by a finite energy}, correlations of local physical excitations decay exponentially at large distances and are therefore short-ranged. This argument, however, does not apply to the topological contribution to the vacuum energy, because it is not associated with propagating physical states. {\it This contribution can remain sensitive to the global structure and large-scale geometry of spacetime even in a gapped theory}. Our proposal is that the observed DE may originate from precisely such a non-local topological contribution.

In the following sections, we develop the framework progressively. First we illustrate, through an example, that topological vacuum structure of gauge theories can give rise to non-local effects. We then turn to the two-dimensional Schwinger model, which provides an exactly solvable realization of this mechanism, before considering a four-dimensional spacetime in which the relevant calculation can also be performed analytically. Finally, we generalize these results to the Friedmann Lemaître Robertson Walker (FLRW) Universe.

\iffalse
In this framework, the Universe will naturally evolve towards the de Sitter geometry, as a result of the presence of topologically nontrivial sectors $|k\ra$ in a time-dependent, expanding universe 
In our proposal, the Dark Energy (DE) is identified with the vacuum energy produced by tunnelling transitions between different topological sectors. The amplitude of DE is associated with the QCD energy scale, while its time variation arises from changes in the topological sectors $|k\ra$  and the corresponding tunnelling transition rates in an expanding universe.

It may be instructive to develop an intuitive picture—presented in Sect.~\ref{intuition}—of the vacuum energy in this framework, formulated through an analogy with condensed matter (CM) systems. This will illustrate, with a simple picture, why the topology of the vacuum in gauge theories is important. Such an intuitive perspective can be helpful in gaining a basic understanding of the nature of DE within the framework proposed in this work.

It is important to emphasize that many key elements of this proposal have been tested in simplified gauge theories, where the computations can be done analytically. The relevant references, along with a list of important technical components used in this construction, are provided in subsections \ref{holonomy} and \ref{sect:EoS}.
\fi
 
\subsection{Non-local effects in gauge theories}\label{intuition}

It may seem counter-intuitive that a local field theory can give rise to genuinely non-local effects. Nevertheless, well-established theoretical and experimental examples exist, such as the screened superconducting realization of the Aharonov--Casher effect \citep{PhysRevD.40.4178}. Consider an electric charge placed inside a cavity surrounded by a superconductor. The mobile charges in the superconductor rearrange so that the electric field is exponentially screened inside the bulk over a short screening length $\lambda$, which plays the role of the gap in QCD DE proposal. A neutral magnetic fluxon propagating through the screened region remains sensitive to the enclosed charge through the quantum phase acquired when it winds around the cavity, even when it travels at distances $r\gg\lambda$ where the local electric field is zero \citep{PhysRevD.40.4178}. The effect is not mediated by a long-range propagating electromagnetic field, but instead reflects global, topological information carried by the gauge configuration. These topological gauge sectors are distinguished by a global winding or flux information that cannot be inferred from the local, exponentially screened electromagnetic field.

In non-Abelian QCD, the analogous structures are the topological sectors $|k\rangle$ of the QCD gauge fields. A QCD topological sector $|k\rangle$ is not a localized region of space, but a class of gauge-field configurations defined over the entire spacetime that share the same integer-valued topological (winding) number $k$. Configurations belonging to different sectors cannot be continuously deformed into one another by ordinary, small gauge transformations; such transitions instead involve non-perturbative configurations such as instantons.

Although the sectors $|k\rangle$ are not associated with physical particles, quantum tunnelling between them modifies the ground-state energy of the theory. The topological structure of the QCD vacuum therefore contributes to the vacuum energy independently of the contribution from propagating degrees of freedom. As a result, the vacuum energy contains two distinct components: one associated with the fields and another associated with the topology of the vacuum.

Now imagine studying the same effect in a time-dependent background. The corresponding topological sectors $|k\ra$ will be modified due to the variation of the external background, a modification which cannot be described in terms of any local dynamical fields. By analogy with the Aharonov-Casher effect, our proposal is that the time evolution of Dark Energy (DE) is governed by the modification of the QCD topological sectors $|k\ra$ as the Universe expands.

\subsection{A mathematically exact analogy}
\label{sec:lessintuition}

The objective of this subsection is to provide a mathematically exact derivation of the dependence of the topological vacuum contribution with the size of the system, thereby making its non-local character explicit. Such a calculation can be carried out in the Schwinger model, namely quantum electrodynamics (QED) in two spacetime dimensions ($1+1$ dimensions) \cite{Urban:2009wb}. Despite its apparent simplicity it shares several important qualitative features with the complicated four-dimensional QCD. In particular, its physical spectrum is characterized by a mass gap, with a single massive bosonic excitation of mass $m_\gamma$, while the gauge vacuum possesses a non-trivial topological structure, with distinct sectors connected by non-perturbative tunnelling processes. The Schwinger model provides a way to study the interplay between a mass gap, vacuum topology, and sensitivity to the large-scale properties of spacetime.

The mass $m_\gamma$ plays a role analogous to that of the strong coupling scale $\Lqcd$ in QCD, while instanton configurations mediate tunnelling between distinct topological sectors. One would expect the physics to become insensitive to the size $\mathbb{L}$ of the system once $\mathbb{L}\gg m_\gamma^{-1}$, since effects associated with massive propagating degrees of freedom should be exponentially suppressed, $\propto \exp(-m_\gamma \mathbb{L})$. Remarkably, this expectation is not realized for the topological contribution: the deviation from the infinite-volume Minkowski result instead scales linearly with $\mathbb{L}^{-1}$. The corresponding corrections were computed explicitly in \cite{Urban:2009wb}, where the topological susceptibility, chiral condensate, and vacuum energy were shown to retain a power-law sensitivity to the size of the system.

Technically, the correction $\propto \mathbb{L}^{-1}$ originates from the zero-mode contribution to the determinant evaluated in the instanton background describing tunnelling between different topological sectors. On a torus of linear size $L$, \cite{Urban:2009wb} find a correction to the topological susceptibility, and hence to the $\theta$-dependent vacuum energy $\Delta\rho_{\rm vac}$, of the form
\be
\Delta\rho_{\rm vac}\propto \frac{1}{m_\gamma L},
\label{eq:Schwinger}
\ee
rather than the exponentially suppressed behaviour $\sim\exp(-m_\gamma L)$. The implication is that tunnelling between topological sectors gives rise to an infrared, non-local sensitivity of the vacuum to the global geometry at distances $L\gg m_\gamma^{-1}$, even though no massless propagating degree of freedom is present. The Schwinger model provides an explicit example demonstrating that a gapped gauge theory can retain sensitivity to arbitrarily large scales through its topological vacuum structure.

\subsection{Generating DE from QCD topological sectors}\label{holonomy}

The generalization to a FLRW Universe remains fundamentally motivated by Zel'dovich's conjecture \cite{Zeldovich:1967gd}, with DE given by Eq.~(\ref{eq:DEdef}), but an exact calculation of the QCD topological contribution in a general FLRW spacetime is not analytically tractable. Such calculations have, however, been carried out in a particular case: a static relativistic hyperbolic spacetime \cite{Zhitnitsky:2013pna,Zhitnitsky:2015dia,Barvinsky:2017lfl}. Our strategy is to use these exact results as a baseline for extending the framework to the FLRW case. We begin by reviewing the corresponding analytical calculations and their main implications.

The relativistic hyperbolic spacetime, $\mathbb{H}^3_{\kappa} \times \mathbb{S}^1_{\kappa^{-1}}$, is characterized by the curvature parameter\footnote{Note that we define $\kappa$ such that $\kappa>0$.} $\kappa$, where $\mathbb{H}^3_{\kappa}$ denotes three-dimensional hyperbolic space, and $\mathbb{S}^1_{\kappa^{-1}}$ is a one-dimensional circle (or ring) with circumference $\kappa^{-1}$. The vacuum energy of the QCD topological sectors associated with this geometry is denoted by $E_{\rm vac}[ \mathbb{H}^3_{\kappa}\times \mathbb{S}^1_{\kappa^{-1}}]$, it is given by \cite{Zhitnitsky:2015dia}:
\be
E_{\rm vac}[ \mathbb{H}^3_{\kappa}\times \mathbb{S}^1_{\kappa^{-1}}] \approx -\Lqcd^4  \left(1-   c_{\kappa} \frac{\kappa}{\Lqcd} \right)
\label{eq:Evac_FLRW}
\ee
where numerical factors are omitted and $c_{\kappa} $ is a positive dimensionless numerical coefficient of order one. 
The overall negative sign in front of Eq (\ref{eq:Evac_FLRW}) is a well-known feature of QCD, originally introduced through the so-called Bag constant, a phenomenological parameter in the MIT bag model\footnote{An intuitive picture is that the true non-perturbative QCD vacuum behaves as a bound state of the gauge theory and therefore lies below the perturbative vacuum in energy. The MIT bag model makes this picture very concrete. It assumes that the physical QCD vacuum outside a hadron has lower energy density than the perturbative vacuum inside the bag.} \citep{Chodos:1974je, Chodos:1974pn}. The negative sign in front of $c_{\kappa}\kappa$, arises from explicit computations in \citep{Zhitnitsky:2015dia}, and can be interpreted as the negative sign encountered in Casimir-type energy calculations.

In comparison, the vacuum energy of flat Minkowski spacetime is given by $E_{\rm vac}[\mathbb{R}^3 \times \mathbb{S}^1]\sim -\Lambda_{\rm QCD}^4$, where $\mathbb{R}^3$ is three-dimensional Euclidean space and $\mathbb{S}^1$ is a one-dimensional circle. Following Eq (\ref{eq:DEdef}), the vacuum energy density $\Delta E_{\rm vac}$ is given by:
\be
\label{vacuum_energy}
&&\Delta E_{\rm vac}\equiv E_{\rm vac}[ \mathbb{H}^3_{\kappa}\times \mathbb{S}^1_{\kappa^{-1}}]
 -E_{\rm vac}[\mathbb{R}^3 \times \mathbb{S}^1] \approx \\
&&    - \left[\Lqcd^4  \left(1-   c_{\kappa} \frac{\kappa}{\Lqcd} \right)
   -  \Lqcd^4\right]\approx c_{\kappa} \kappa\cdot \Lqcd^3, \nonumber
\ee
which is positive, as expected. Eq (\ref{vacuum_energy}) reveals an infrared sensitivity to the spacetime curvature, analogous to the infrared sensitivity of the Schwinger model exhibited in Eq.~(\ref{eq:Schwinger}). This is a non-trivial result, since it is precisely this infrared dependence that sets the characteristic amplitude of the DE contribution. The ordinary local contribution from propagating particles and fields scales as $\Lambda_{\rm QCD}^4$ and is removed by the Minkowski subtraction. More generally, after integrating out a massive sector, local QFT in curved spacetime typically generates corrections of order $\sim R\,\Lambda_{\rm QCD}^2\sim \kappa^2\,\Lambda^2_{\rm QCD}$. Such a quadratic contribution would be far too small to account for the observed dark-energy density, illustrating the difficulty of explaining DE through conventional local propagating degrees of freedom. By contrast, the topological sector of the QCD vacuum generates a contribution $\Delta\rho_{\rm vac}\propto \kappa\,\Lambda_{\rm QCD}^3$, whose magnitude is considerably larger than the quadratic contribution, offering a possible solution to the DE amplitude problem.

The generalization of the framework to a FLRW Universe proceeds in two steps. We first consider the de Sitter case, corresponding to the asymptotic future evolution of the Universe, and then turn in Section~\ref{DESI} to the more general FLRW spacetime.

At present, however, a first-principles calculation cannot be carried out explicitly in de Sitter spacetime \citep{Zhitnitsky:2015dia}. One of the main obstacles is that monopole configurations with non-trivial holonomy are not known in this geometry. We therefore assume that the positivity of $\Delta E_{\rm vac}$ and its infrared behavior persist in the de Sitter case, as expected for a non-trivial holonomy configuration. Following \cite{Barvinsky:2017lfl}, we further conjecture that the Hubble constant replaces the quantity $\kappa$ in Eq.~(\ref{vacuum_energy}), with the corresponding dimensionless coefficient $c_{\kappa}\sim1$ replaced by another coefficient $c_H\sim1$. This constitutes a central assumption of the present work and is motivated by the exact calculations and theoretical arguments developed in \citep{Zhitnitsky:2015dia,Barvinsky:2017lfl}. Within this framework, the DE density in the vicinity of the de Sitter state is therefore given by
\be
\rho_{\rm DE}=\Delta E_{\rm vac}\sim H\Lambda_{\rm QCD}^3.
\label{eq:deSitter}
\ee
Further technical details of the QCD-induced dark-energy mechanism are provided in Appendix~\ref{sect:technique}.

\subsection{QCD-induced DE in de Sitter spacetime and its equation of state}\label{sect:EoS}

Since Eq.~(\ref{eq:deSitter}) is in principle only valid in the de Sitter limit, we first consider the asymptotic regime in which the Hubble parameter approaches a constant value, denoted by $\overline{H}$. The scale factor then evolves as $a(t)\propto \exp(\overline{H}t)$, and the DE density and asymptotic Hubble parameter satisfy
\be
\label{delta1}
\overline{H}^2=\frac{8\pi G }{3}  \rho_{\rm DE}, ~~  \rho_{\rm DE} & \approx& c_H \Lqcd^3 \overline{H}, ~~G\equiv M_{PL}^{-2} \nonumber \\
 \overline{H}=c_H\frac{8\pi \Lqcd^3}{3M_{PL}^2}, ~  \rho_{\rm DE}   &\approx& c_H^2 \frac{8\pi \Lqcd^6}{3M_{PL}^2}, 
\ee
The resulting scaling is identical to that of Zel'dovich's original estimate, Eq.~(\ref{eq:Zeldo}), with the proton mass $m_p$ replaced by the QCD scale $\Lqcd$. Taking $\Lqcd\approx100,{\rm MeV}$ and assuming that the dimensionless coefficient $c_H$ is approximately constant\footnote{No generality is lost by adopting $\Lqcd\approx100\,{\rm MeV}$, since the dimensionless coefficient $c_H$ may be redefined to absorb numerical factors entering the calculation. In particular, the small QCD factor $\propto m_q/\Lqcd\approx0.05$ associated with the tunnelling transitions can be incorporated into $c_H$; here we adopt a representative light-quark mass $m_q\approx5\,{\rm MeV}$.}, one obtains
\be
\label{numerics}
\overline{H}&=&c_H\frac{8\pi \Lqcd^3}{3M_{PL}^2}\approx \bar{c}_{{H}} \cdot 2.8\cdot 10^{-33} eV, \\
&\simeq & \bar{c}_{H}\cdot 130\,{\rm km/s/Mpc}\nonumber\\
t_0&\equiv&  {\overline{H}}^{-1}=\frac{7.3}{\bar{c}_{{H}}} {\rm  Gyr}, ~~ c_H\equiv \bar{c}_{{H}} \left(\frac{m_q}{\Lqcd}\right)\approx 0.05 \bar{c}_{{H}}, \nonumber\\
\rho_{\rm DE} &=&c_H \Lqcd^3 \overline{H} \approx \bar{c}_{{H}}^2 \left(3.4\cdot 10^{-3} {\rm eV} \right)^4, ~~\Omega_{\rm DE}\sim 1\nonumber
\ee
For an order-unity value of $\bar c_H$, these estimates lie close to the observed values. This is particularly remarkable in view of the roughly $120$ orders of magnitude separating the quantum-field-theory estimate of the vacuum energy from the observed DE density.
In our framework, the relevant energy scale emerges directly from $\Lqcd$, without introducing an additional finely tuned mass scale, suggesting a possible physical origin for both the small magnitude of DE and the weakness of its associated interaction. Whether the same framework can address why dark energy becomes cosmologically important only at late times, will be explored in Section~\ref{DESI}.

The equation-of-state parameter $\omega_{\rm DE}$ is a central quantity in dark-energy phenomenology and one of the primary observables constrained by cosmological data; it is therefore important to determine its effective value within the QCD-DE framework. Although DE in this framework is not associated with any propagating degree of freedom, one can still define an effective equation of state (EoS) that reproduces the same background evolution. This effective EoS should not be interpreted as the physical equation of state of an underlying fluid; rather, it provides a phenomenological description of how an observer unaware of the QCD origin of dark energy would infer its behavior from the cosmological expansion history. Using standard thermodynamic principles:

 \be
\label{thermodynamics}
dF=TdS-PdV, ~~~~ P=-\left.\frac{\partial F}{\partial V}\right|_S,
\ee
to arrive 
\be
\label{P-rho}
P&=&-\frac{\partial F}{\partial V}=+  \Lqcd^4  \left(1-  c_H \frac{ \overline{H} }{\Lqcd} \right)\nonumber\\
\rho&=&\frac{F}{V}= -   \Lqcd^4  \left(1-   c_H\frac{ \overline{H} }{\Lqcd}  \right).   
\ee

After subtraction of the Minkowski contributions, we arrive to the following expressions for the vacuum pressure and energy:

\be
\label{eq:P-rho-subtracted}
P_{DE} = -  c_H \Lqcd^3 \overline{H}, ~~~
\rho_{DE}=  +   c_H \Lqcd^3 \overline{H}     
\ee
such that 
the equation  of state   assumes the  form 
\be
\label{EoS}
\omega_{\rm DE} &=& \frac{ P_{DE}}{ \rho_{DE}}= -1  , ~~  {  a}(t)\propto \exp (\overline{H} t),  
\ee
which is precisely the expected EoS in de Sitter. We refer the reader to Appendix~\ref{misuse} for further discussion of the physical interpretation of Eq.~(\ref{eq:P-rho-subtracted}).

In the next section, we generalize the framework to a FLRW Universe.

\section{DESI results and deviation from exact de Sitter behaviour}\label{DESI}
  
The DESI results \citep{2025arXiv250314738D} suggest not only that the DE EoS parameter $\omega_{\rm DE}$ may differ from $-1$, but also that it may evolve with time and even cross the $\omega_{\rm DE}=-1$ boundary. The formalism developed in Section \ref{sec:topology}, however, applies only in the de Sitter limit with constant $\bar{H}$. Extending this framework to finite redshift requires understanding how a time-dependent cosmological background modifies the QCD vacuum calculations. This is the purpose of the present section.

\subsection{The adiabatic condition}
\label{sect:beta}

When the QCD-induced dark-energy mechanism operates in a background characterized by a constant $\overline{H}$, the system does not depart from a pure de Sitter geometry and Eq.(\ref{eq:deSitter}) applies with $H=\overline{H}$. When the Hubble parameter varies with time, Eq.(\ref{eq:deSitter}) will be suppressed. The physical origin of this suppression can be understood from Eq.~(\ref{vacuum_energy}): the key ingredient in that analysis is the pre-factor $\kappa/\Lqcd$, which originates from the zero-mode determinant entering the tunnelling transition rate. The corresponding calculation necessarily involves integration over collective coordinates, including spatial and temporal coordinates, extending over distances of order $\kappa^{-1}$. Suppose now that a time-dependent perturbation\footnote{For instance, a change in the expansion rate of the Universe $H=\overline{H}+\delta H$.} is introduced with a characteristic frequency $\omega \gg \kappa$, so that the timescale of the perturbation is much shorter than the characteristic size of the system, $\kappa^{-1}$. In this regime, an additional suppression factor proportional to $(\kappa/\omega)\ll 1$ arises. Physically, this occurs because the zero mode is no longer an exact zero mode and therefore cannot probe distances as large as $\kappa^{-1}$. The effective range of the corresponding collective coordinate is consequently reduced by a factor of order $\kappa/\omega$. Equivalently, the integration over the collective coordinate can no longer extend to its maximal scale $\kappa^{-1}$, but is instead effectively truncated at the shorter scale $\omega^{-1}$. This behaviour reflects the intrinsic sensitivity of the zero mode, and hence of the tunnelling contribution, to the largest available length scales. As we will see below, these arguments are supported by explicit computations in the 2-dimensional Schwinger model.

The origin of the suppression can be used to generalize the de Sitter limit $\rho_{\rm DE}=c_H\Lqcd^3\overline{H}$ to the case of a time-dependent Hubble parameter $H(t)$. Such a generalization is expected to remain valid provided that the condition
\be
\label{eq:adiabatic}
\frac{|\dot{H}|}{H} \ll \overline{H}
\ee
is satisfied. This condition is the cosmological analogue of the requirement $\omega\ll\kappa$ discussed in the previous paragraph and corresponds to a background that remains sufficiently close to de Sitter. Indeed, when the Hubble parameter varies with time, the characteristic frequency of the perturbation should be identified as
$\omega\rightarrow |\dot{H}|/H$, while the parameter $\kappa$, which sets the inverse size of the system, is identified with the constant de Sitter expansion rate, $\kappa\rightarrow\overline{H}$. Equation~(\ref{eq:adiabatic}) will therefore be referred to as the \emph{adiabatic condition}.

Interestingly, the two-dimensional Schwinger model on a torus, described in Section~\ref{sec:lessintuition}, provides an independent motivation for the adiabatic condition (see Appendix \ref{2d_QED}). When the size of the system, $L$, is allowed to vary with time, thereby mimicking an expanding Universe, the departure from Eq.~(\ref{eq:Schwinger}) takes the form:
\be
\frac{\Delta \rho_{\rm vac}}{\Lambda^4_{\rm QCD}}
\propto
\frac{\overline{H}/\Lqcd}
{1+|\dot{H}|/(H\overline{H})}.
\label{eq:schwinger_adia}
\ee
The de Sitter scaling of Eq.(\ref{eq:P-rho-subtracted}) is recovered from Eq.~(\ref{eq:schwinger_adia}) when $\dot H\ll H\overline{H}$, i.e. in the adiabatic regime Eq.(\ref{eq:adiabatic}). When the adiabatic condition is violated. Eq.(\ref{eq:schwinger_adia}) shows the suppression factor $\kappa/\omega \ll 1$ as discussed above. Although the adiabatic condition has not yet been derived from first principles in four-dimensional QCD, the Schwinger-model and the relativistic hyperbolic spacetime results provide a physically motivated prescription for describing departures from the de Sitter limit, which can then be used phenomenologically in an FLRW background.

\subsection{The de Sitter limit}
\label{sect:desitter}

For a spatially flat Universe, the Friedmann equation can be written as
\be
1=\Omega_{\rm m}(z)+\Omega_{\rm r}(z)+\Omega_{\rm DE}(z),
\ee
where $\Omega_{\rm m}(z)$, $\Omega_{\rm r}(z)$, and $\Omega_{\rm DE}(z)$ denote the matter, radiation, and DE density parameters at redshift $z$, respectively. Close to the de Sitter limit, we write the Hubble parameter as $H=\overline{H}+\delta H$. In the adiabatic regime,
\be
\overline{H}=
H\left[
1+\mathcal{O}\left(
\frac{|\dot H|}{H^2},
\frac{\delta H}{H}
\right)
\right].
\ee
It is justified, to leading order, to promote the de Sitter expression
$\rho_{\rm DE}=c_H\Lqcd^3\overline{H}$ to the time-dependent form
$\rho_{\rm DE}=c_H\Lqcd^3 H$.

We then rewrite the density parameters as a function of the scale factor $a\propto (1+z)^{-1}$ and the Hubble parameter $H(z)$:

\be
\Omega_{\rm m}&=&\frac{\rho_{\rm m}}{\rho_{\rm crit}}=\frac{\rho_{\rm m,i}\left( a_i/a\right)^3}{\rho_{\rm crit,i}\left( H/H_i\right)^2}=\left(\frac{H_i}{H}\right)^2\Omega_{\rm m,i}\left(\frac{a_i}{a}\right)^3 \nonumber\cr
\Omega_{\rm r}&=&\frac{\rho_{\rm r}}{\rho_{\rm crit}}=\frac{\rho_{\rm r,i}\left( a_i/a\right)^4}{\rho_{\rm crit,i}\left( H/H_i\right)^2}=\left(\frac{H_i}{H}\right)^2\Omega_{\rm r,i}\left(\frac{a_i}{a}\right)^4\nonumber\cr
\Omega_{\rm DE}&=&\frac{\rho_{\rm DE}}{\rho_{\rm crit}}=\frac{c_H\Lambda_{\rm QCD}^3 H}{\frac{3H^2}{8\pi G}}=\frac{8\pi G}{3}  \Lambda_{\rm QCD}^3 \frac{c_H}{H}=  \frac{\overline{H}}{H},
\ee
where $\overline{H} \equiv c_H\frac{8\pi G}{3}\Lambda_{\rm QCD}^3$ includes the coefficient $c_H$ to be consistent with (\ref{delta1}) and (\ref{numerics}). The quantities $(H_i, a_i)$ are specified at an arbitrary reference time $i$. We choose to anchor the Friedmann equation at an early epoch when the classical Friedmann equation holds with DE negligible. The Friedmann equation can now be rewritten as:

\be
H^2-\overline{H} H-H_{\rm i}^2\left[ \Omega_{\rm m,i}\left(\frac{a_i}{a}\right)^3+\Omega_{\rm r,i}\left(\frac{a_i}{a}\right)^4\right]=0
\label{eq:QCDfriedmann}
\ee
This is a second order polynomial in $H$, whose physical solution is given by:
\be
H(a) = \frac{ \overline{H}}{2}\left(1 +\sqrt{1+B \left(\frac{a_i}{a}\right)^3+C \left(\frac{a_i}{a}\right)^4}\right)
\label{eq:H},
\ee
where $B\equiv 4\left(\frac{H_i}{\overline{H}}\right)^2\Omega_{m,i}$ and $C\equiv 4\left(\frac{H_i}{\overline{H}}\right)^2\Omega_{r,i}$. It is straightforward to verify that Eq.~(\ref{eq:H}) reduces to the classical Friedmann equation with only matter and radiation   when $\overline{H} \rightarrow 0$ when $\sqrt{B}$ and $\sqrt{C}$ are both proportional to $\overline{H}^{-1}$ to cancel $\overline{H}$ in front. One can also verify that $H(a\rightarrow\infty)\rightarrow \overline{H}$, which confirms the physical meaning of $\overline{H}$ as an  asymptotic value for the Hubble constant in far future.

We want to compute the solution of Eq.~(\ref{eq:H}) and the DE EoS when $a\gg 1$, i.e. when the system is not in the exact de Sitter state, but approaching  it. In this limit, the radiation term can be neglected and we can choose the reference time $i$ to be today, allowing us to express the departure from de Sitter as the present epoch is approached from the future. Taking \(c_H\) to be constant, Eq.~(\ref{eq:H}) can be written as:

\be
\label{eq:solution}
H=\frac{ \overline{H}}{2}+\frac{ \overline{H}}{2}\sqrt{1+4 \Omega_{M,0} \left(\frac{H_0^2}{ \overline{H}^2}\right) \left(\frac{a_0}{a(t)}\right)^3},
\ee
In the de Sitter limit, when $a\rightarrow \infty$, we have $H\rightarrow \overline{H}$. Differentiating with respect to time $t$, we get:
\be
\label{eq:H-dot}
\dot{H}=\frac{\partial H}{\partial a} {H} a=-\frac{3  \overline{H} H\Omega_{M,0} \left(\frac{H_0^2}{ \overline{H}^2}\right)\left(\frac{a_0}{a(t)}\right)^3}{\sqrt{1+4 \Omega_{M,0} \left(\frac{H_0^2}{ \overline{H}^2}\right) \left(\frac{a_0}{a(t)}\right)^3}}
\ee 
where we  use $\dot{a}= {H} a$. Using the acceleration equation and the DE equation of state $P_{\rm DE}=\omega_{\rm DE}\rho_{\rm DE}$, we can write $\dot H$ as a sum over the mass densities:

\be
\dot{H} = -4\pi G \left( \rho_m + \rho_{\rm DE}(1 + \omega_{\rm DE}) \right)
\label{eq:acc}
\ee

Combining the DE mass density $\rho_{\rm DE}\equiv c_H\Lambda_{\rm QCD}^3 H$ with Eq.(\ref{eq:solution}), we get:

\be
\label{eq:rho2}
\rho_{\rm DE} =  \frac{3\overline{H}H}{8\pi G} = \frac{3 \overline{H}^2}{16\pi G} \left[ 1+\sqrt{1+4 x(t)} \right],
\ee

where we have defined $x(t)$:

\be
\label{eq:x}
x(t)&\equiv &\Omega_{M,0} \left(\frac{H_0^2}{ \overline{H}^2}\right)\left(\frac{a_0}{a(t)}\right)^3\nonumber\cr
&=&\Omega_{M,0} \left(\frac{H_0^2}{ \overline{H}^2}\right) (1+z)^3.~~
\ee
 
Similarly, we can express the matter density as:

\be
\rho_{\rm m}=\frac{3 \overline{H}^2}{8\pi G} x(t)
\label{eq:matter}
\ee

Combining Eqs. (\ref{eq:acc}, \ref{eq:rho2}, \ref{eq:matter}) we obtain the DE EoS parameter that is asymptotically justified for $x(t)\ll 1$, i.e. as long as DE is the dominant contributor to the Universe expansion:
 
\be
\label{eq:w}
\omega_{\rm DE}+1=\frac{ \rho_{\rm DE} +P_{\rm DE}}{ \rho_{\rm DE}}=\frac{\left[ \frac{2x(t)}{\sqrt{1+4 x(t)}} \right]}{\left[ 1+\sqrt{1+4 x(t)} \right]},
\ee

One can see that $\omega_{\rm DE} \rightarrow -1$ as $x(t) \rightarrow 0$, recovering our earlier Eq. (\ref{EoS}) for the de Sitter state. Interestingly, a formal extrapolation of Eq.(\ref{eq:w}) towards $x(t_0) \sim 1$, gives $\omega_{\rm DE} \simeq -0.7$, illustrating two important consequences of our framework, qualitatively consistent with the behaviour suggested by DESI: (1) the equation-of-state parameter $\omega_{\rm DE}$ is time-dependent, and (2) it is possible to have $\omega_{\rm DE} > -1$ at the present epoch.

Substituting the corresponding expressions for $\dot{H}$ and $H^2$ in terms of $\rho_{\rm DE}$, $P_{\rm DE}$, and $\rho_{\rm M}$, the following condition must be satisfied:

\be
\label{eq:adiabatic_EoS}
\frac{3}{2}\left(1+\frac{\omega_{\rm DE} \rho_{\rm DE}}{\rho_{\rm DE}+\rho_{\rm M}}\right)\ll \frac{\overline{H}}{H},
\ee
which can only be justified when $\rho_{\rm M}\ll\rho_{\rm DE} $
(in which case  $\omega_{\rm DE}\approx -1$). The condition (\ref{eq:adiabatic_EoS}) is the adiabatic condition applied to the EoS parameter. It is only marginally satisfied today at $z=0$, and it is strongly violated at $z\gtrsim 1$.

\subsection{Beyond de Sitter}

The phenomenological extension of the Schwinger-model result to a time-dependent background, Eq.~(\ref{eq:schwinger_adia}), provides some guidance on the behaviour away from the de Sitter limit. In the standard cosmological model, $\dot H\rightarrow -\infty$ as $a\rightarrow 0$, implying that the topological contribution to $\rho_{\rm DE}$ is strongly suppressed and tends to zero near the Big Bang\footnote{This differs from $\Lambda$CDM, where $\rho_{\rm DE}$ remains constant. The distinction is irrelevant at early times, however, since dark energy is then overwhelmingly subdominant to radiation and matter.}. The precise rate and redshift dependence of this suppression in a general FLRW background are not known. Nevertheless, Eq.~(\ref{eq:schwinger_adia}) indicates that the DE contribution becomes significant only as the evolution approaches the adiabatic regime. As discussed in Section~\ref{sect:desitter}, and specifically Eq.(\ref{eq:adiabatic_EoS}), this can occur only at relatively late times, $z\lesssim 1$.

In order to implement this suppression we introduce a dimensionless time-dependent function $\beta(t)$ that approaches unity in the asymptotic de Sitter regime,
$\beta(z\rightarrow -1)\rightarrow 1$, while tending to zero at sufficiently high redshift, $\beta(z\gtrsim 1)\rightarrow 0$. The DE density is then parameterized as
\be
\label{eq:extrapolation}
\rho_{\rm DE}=\beta(t)\,c_H\Lqcd^3 H,
\qquad
\beta(t)\in(0,1).
\ee
The function $\beta(t)$ controls the gradual activation of the DE contribution as the Universe approaches the adiabatic regime. Its introduction is not an ad hoc modification of the model, but rather a phenomenological implementation of the suppression implied by the adiabatic condition near the de Sitter limit.

In principle, the coefficient $\beta(t)c_H$ could be computed from first principles for a specified spacetime geometry at each time $t$, together with the corresponding subtraction of the Minkowski-space contribution, in direct analogy with the procedure leading to Eq.~(\ref{P-rho}) in the pure de Sitter case with constant $\overline{H}$.
\begin{figure}
\centering
\hspace*{-0.5cm}\includegraphics[width=0.47\textwidth]{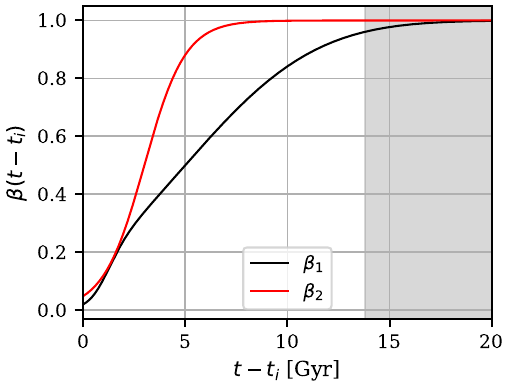}
\caption{The two parameterizations of $\beta(t)$ defined in Eq.~(\ref{eq:betas}). The time $t_i=0.0465;{\rm Gyr}$ corresponds to the anchor point at which the solution is matched to the standard Friedmann evolution. Since this time lies very close to the Big Bang, $t-t_i=0$ may be regarded as the effective beginning of the cosmological evolution considered here. The grey region beyond $t-t_i=13.78;{\rm Gyr}$ corresponds to the future.}
\label{fig:beta}
\end{figure}
As we saw in Section \ref{holonomy}, such a first-principles calculation for QCD in a general FLRW spacetime is not currently feasible, even with modern lattice-QCD techniques. In what follows, we consequently treat $\beta(t)$ as a phenomenological parameterized function.

For $z\geq0$, the evolution of $a(t)$ must be modified to include the time dependence of $\beta(t)$. Substituting Eq.~(\ref{eq:extrapolation}) into the Friedmann equation yields the following differential equation for $a(\tau)$:
\begin{eqnarray}
\frac{da}{d\tau}&=&\beta(\tau)\frac{a}{2}\times \nonumber\\
&&\left(
1+\sqrt{
1+\frac{B}{\beta^2(\tau)}
\left(\frac{a_i}{a}\right)^3
+\frac{C}{\beta^2(\tau)}
\left(\frac{a_i}{a}\right)^4
}
\right),
\label{eq:aoftau}
\end{eqnarray}
where we have introduced the dimensionless time variable $\tau\equiv (\frac{8\pi G}{3}\Lambda_{\rm QCD}^3)c_H t$. Eq.(\ref{eq:aoftau}) generalizes Eq.(\ref{eq:H}) to the case of a time-dependent $\beta(t)$ and can be solved numerically once $\beta(t)$ is specified. In deriving Eq.(\ref{eq:aoftau}), the explicit time dependence has been factored out of the coefficients $B$ and $C$, so that

$$
\overline{H}
=
c_H\left(\frac{8\pi G}{3}\Lambda_{\rm QCD}^3\right)
$$

remains constant and corresponds to its asymptotic de Sitter value, reached in the far future when $\beta\rightarrow1$ (see Section~\ref{sect:desitter}).

To illustrate the cosmological impact of a time-dependent $\beta(t)$, and to compare it with the standard $\Lambda$CDM evolution, we consider two qualitatively different functional forms: $\beta_1(t)$, for which the QCD-induced DE contribution turns on gradually over cosmic time, and $\beta_2(t)$, which produces a much sharper transition:
\be
\beta_1(t)&=&\frac{1}{2}\left[1+\operatorname{erf}\left(\frac{t-t_1}{\Delta t_1}\right)\right]
\frac{1}{1+\exp\left[-(t-t_2)/\Delta t_2\right]}
\cr
\beta_2(t)&=&\frac{1}{1+\exp\left[-(t-t_3)/\Delta t_3\right]},
\label{eq:betas}
\ee
where we choose $t_1=\Delta t_1=5;{\rm Gyr}$, $t_2=1;{\rm Gyr}$, $\Delta t_2=0.5;{\rm Gyr}$, $t_3=3;{\rm Gyr}$, and $\Delta t_3=1;{\rm Gyr}$. These values are not associated with any specific physical transition and are not obtained from a fit to the data. They are chosen solely to illustrate two different timescales over which the QCD-induced DE contribution may become cosmologically relevant. The two functions are shown in Figure~\ref{fig:beta}.

We then solve Eq.~(\ref{eq:aoftau}) numerically, adopting $\overline{H}=50\,{\rm km\,s^{-1}\,Mpc^{-1}}$. The initial condition is imposed at $z_i=50$, using the Planck cosmology values $\Omega_{\rm m,i}\simeq0.982$ and $\Omega_{\rm r,i}\simeq0.018$. At this redshift, the DE contribution is negligible, with $\Omega_{\rm DE,i}\simeq1.65\times10^{-5}\ll1$, as required. The corresponding Hubble parameter is $H_i\simeq1.38\times10^5\;{\rm km\,s^{-1}\,Mpc^{-1}}$. Since $H_i\gg\overline{H}$ and $\beta(t)\ll1$ at high redshift, this naturally implies $B,C\gg1$, without requiring any fine-tuning of $\overline{H}$. We take $a_i=1/(1+z_i)\simeq0.0196$ at the initial time $\tau_i$, corresponding to $t_i\simeq0.0465;{\rm Gyr}$.

\begin{figure*}
  \centering
  \includegraphics[width=0.47\textwidth]{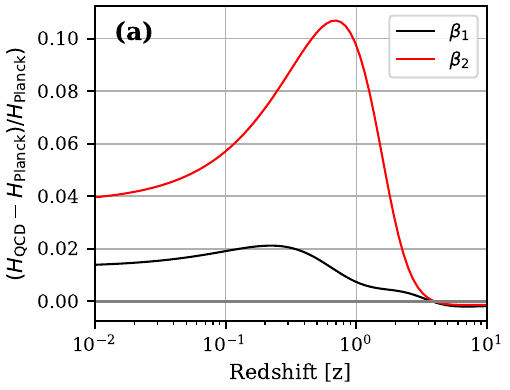}
  \hfill
  \includegraphics[width=0.47\textwidth]{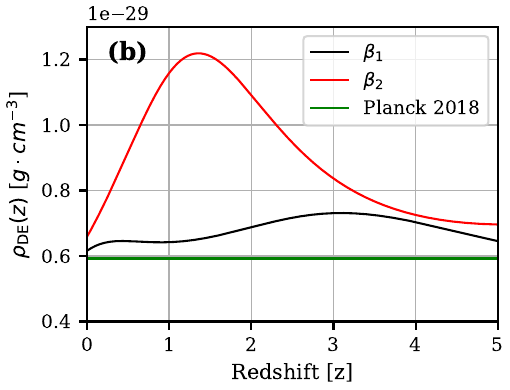}
  \vskip\baselineskip
  \includegraphics[width=0.47\textwidth]{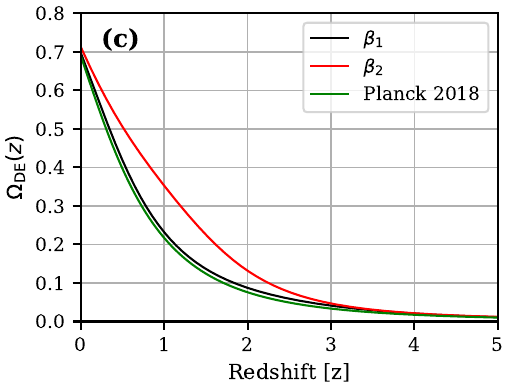}
  \hfill
  \includegraphics[width=0.47\textwidth]{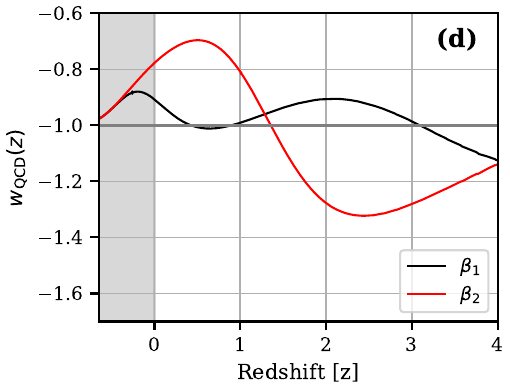}
  \caption{Plots comparing various quantities in our framework to the Planck2018 cosmology as a function of redshift. Black and red curves correspond respectively to the activation functions $\beta_1(t)$ and $\beta_2(t)$. When present, the green curve corresponds to the Planck 2018 cosmology. Panel (a): fractional difference of the Hubble parameter, panel (b): DE mass density $\rho_{\rm DE}$, panel (c): DE dimensionless density parameter, panel (d): DE equation of state parameter $\omega_{\rm DE}$. The grey area for $z<0$ corresponds to the future.}
  \label{fig:QCD_cosmo}
\end{figure*}

Figure~\ref{fig:QCD_cosmo} shows the resulting evolution of the Hubble parameter $H(z)$, the DE density $\rho_{\rm DE}(z)$, the density parameter $\Omega_{\rm DE}(z)$, and the DE equation of state $w_{\rm QCD}(z)$, together with their corresponding $\Lambda$CDM predictions. Panels (a)-(c) illustrate how different histories of QCD DE activation affect key cosmological observables. Panel (d) shows that the effective DE equation-of-state parameter can evolve with time and may cross the $w=-1$ boundary, even multiple times. Figure~\ref{fig:QCD_cosmo} summarizes the main result of this paper: the topological QCD vacuum contribution generates a DE density of the observed order of magnitude, while its effective equation of state can evolve and exhibit phantom-boundary crossings. It illustrates how our realization of Zel'dovich's proposal provides a common physical framework for addressing both the amplitude and the possible time evolution of dark energy without introducing new fundamental fields. Our framework is directly testable cosmologically: CMB anisotropies, BAO, and SNIa measurements can jointly constrain the same function $\beta(t)$, since a single activation history must simultaneously reproduce all of these observables (see companion paper \cite{2026arXiv260620036L}).

Panel (d) also shows that, in the future ($z<0$), $\omega_{\rm DE}$ asymptotically approaches $-1$, in agreement with the de Sitter limit derived in Section~\ref{sect:desitter}. This is a generic prediction of our framework: the Universe evolves toward an asymptotic de Sitter phase, while at the present epoch one expects $\omega_{\rm DE}>-1$. Although this future behaviour cannot be tested directly, it has observable consequences at $z\geq0$, where the evolution of $\omega_{\rm DE}(z)$ is determined by the activation history encoded in $\beta(t)$. This connection follows directly from the relation $\rho_{\rm DE}(t)\propto\beta(t)H(t)$ during the epoch in which dark energy becomes cosmologically important. The predicted relation between $\omega_{\rm DE}(z)$ and the expansion history $H(z)$ provides an indirect test of the asymptotic de Sitter behaviour of the model, as illustrated by panels (a) and (d). Testing this relation requires independent constraints on both $H(z)$ and $\omega_{\rm DE}(z)$, which will improve with future DESI data and Stage-IV surveys such as Euclid, Roman and Rubin.

\section{Conclusion}
\label{conclusion}

Building on earlier work \citep{Zhitnitsky:2013pna,Zhitnitsky:2015dia,Barvinsky:2017lfl}, we have presented a physically motivated model of DE rooted in the topological structure of the QCD vacuum. The motivation is strengthened by recent DESI results \citep{2025arXiv250314738D}, which provide significant evidence for an evolving DE component. In this framework, DE is identified with the residual vacuum energy associated with tunnelling transitions between distinct topological sectors, after subtracting the much larger contribution computed in Minkowski spacetime, following the spirit of Zel'dovich's original proposal \citep{Zeldovich:1967gd}. No new fundamental fields are introduced, and the relevant physical scale is set by a single QCD parameter, $\Lqcd$. The framework is based entirely on the established topological properties of the vacuum in non-Abelian gauge theories, whose possible cosmological consequences have received comparatively little attention.

Our so-called QCD-DE scenario provides a physical framework for addressing the vast separation between particle-physics and cosmological energy scales, and the weakness of the DE interaction. In this framework, the characteristic timescale at which $\rho_{\rm DE}$ becomes dynamically important, is
$t_0=\overline{H}^{-1}\sim 10\,{\rm Gyr}$. Since $\overline{H}$ is itself determined by $\Lqcd$, the observed cosmological timescale emerges directly from the QCD scale.

The central new ingredient introduced in this work is the extension of the earlier de Sitter formulation \citep{Zhitnitsky:2013pna,Zhitnitsky:2015dia,Barvinsky:2017lfl} to backgrounds that depart from the pure de Sitter limit, thereby allowing the model to be confronted with cosmological observations. The main results can be summarized as follows:

1. The DE EoS parameter asymptotically approaches $\omega_{\rm DE}=-1$ in the future, as the Universe evolves toward the de Sitter regime.

2. The resulting Friedmann solutions allow $\omega_{\rm DE}$ to lie either above or below $-1$ and to cross the $w=-1$ boundary, potentially more than once during the cosmological evolution. Such behaviour is qualitatively consistent with the type of evolution suggested by recent DESI analyses.

3. In the vicinity of the de Sitter regime, where dark energy dominates the expansion and $\rho_{\rm DE}\propto\beta H$, the model predicts a direct relation between the evolution of $\omega_{\rm DE}(z)$ and the Hubble parameter $H(z)$. This correlation provides an additional and distinctive observational test of the framework.

4. For $z\geq0$, the value and redshift dependence of $\omega_{\rm DE}(z)$ are governed by an activation function $\beta(t)$, which describes the emergence of the QCD-induced vacuum contribution as the Universe approaches the adiabatic, de Sitter regime. Although $\beta(t)$ cannot presently be computed from first principles in a general FLRW background, it affects the expansion history and hence cosmological distance and timescale observables. The framework can therefore be tested with standard candles, standard rulers, and CMB measurements.

We refer the reader to \cite{2026arXiv260620036L} for the first comparison of this model with CMB, BAO, and SNIa data. That analysis shows that QCD-DE provides a good fit to the observations and can reproduce the late-time DE evolution favoured by DESI. The data show a mild preference for QCD-DE over $\Lambda$CDM, while the model is substantially favoured relative to the $w_0w_a$CDM parametrization.

We conclude with a few remarks on possible future developments of this framework:

1. Topological toy models of dark energy, such as the two-dimensional Schwinger model, may provide physically motivated functional forms for the activation function $\beta(t)$. For example, Eq.~(\ref{eq:schwinger_adia}) suggests that $\beta(t)$ could be replaced by a one-parameter function directly related to the departure from the adiabatic regime. Exploring such a construction will be the subject of future work.

2. Further theoretical work is needed to establish the adiabatic condition from first principles in a general FLRW background and, ultimately, in the presence of cosmological inhomogeneities. This may be particularly relevant if the activation of DE is not perfectly homogeneous. Matter-density fluctuations modify the local expansion rate and could therefore lead to spatial variations in the degree to which the adiabatic condition is satisfied. In this case, the QCD-induced DE contribution would itself develop spatial fluctuations correlated with the matter distribution. Such an effect could mimic an effective coupling between dark energy and dark matter, even in the absence of any direct microscopic interaction between the two components.

3. de Sitter-like phases appear twice in the standard cosmological history: during primordial inflation and again during the present DE-dominated epoch. The mechanism explored here may also offer insight into the origin of inflation. In the speculative scenario proposed in \cite{Zhitnitsky:2013pna,Barvinsky:2017lfl}, the vacuum energy driving inflation could arise from tunnelling transitions in an as-yet unidentified strongly coupled gauge theory, in close analogy with the QCD mechanism considered here for dark energy. Although the corresponding gauge theory is unknown, the characteristic energy scale, playing a role analogous to $\Lqcd$, and the expected number of e-foldings can nevertheless be estimated within this framework \citep{Zhitnitsky:2013pna,Barvinsky:2017lfl}.

4. The QCD-induced DE framework has also been argued to generate cosmological magnetic fields with amplitudes of order $10^{-10}\,{\rm G}$ and correlation lengths comparable to the size of the observable Universe \citep{Zhitnitsky:2019ijg}. Observational indications of magnetic fields coherent over very large cosmological scales have also been reported; see \cite{Durrer:2013pga} for a review. This provides another potentially observable consequence of the underlying topological vacuum dynamics.

5. Finally, some of the key physical ideas underlying this framework, in particular, vacuum-energy contributions associated with tunnelling between topological sectors, may, in principle, be tested in laboratory experiments. A possible avenue is the search for an additional contribution to the Casimir energy in Maxwell theory, as proposed in \cite{Cao:2013na,Cao:2015uza,Yao:2016bps}. This so-called Topological Casimir Effect originates from non-dispersive tunnelling contributions rather than from the conventional quantum fluctuations of propagating photons. Although such a contribution has not yet been observed, its detection would provide direct experimental evidence for the type of topological vacuum physics invoked in the present DE scenario.

\section*{Acknowledgements}
We are very grateful to Robert Brandenberger for many stimulating and useful discussions during his visit to UBC. We also thank him for initiating and motivating this project, and for encouraging us to bring it to completion. LVW thank Benoite Pfeiffer for a careful reading of the manuscript. The authors thank ChatGPT5.6 for comprehensive language editing to improve the academic English phrasing of the entire manuscript. All scientific concepts, calculations, and final edits remain the sole work and responsibility of the human authors. This research was supported in part by the Natural Sciences and Engineering Research Council of Canada.

\appendix

\section{On the nature of long-range order in QCD}
\label{sect:technique}

In this Appendix, we elaborate on several technical aspects of the discussion introduced in Section~\ref{holonomy}, with the aim of clarifying how long-range order can emerge in a theory with a mass gap.

1. The long-range order in a gapped theory manifests itself through a non-dispersive contribution to the topological susceptibility,

$$
\chi=\chi_{\rm non-dispersive}+\chi_{\rm dispersive}.
$$

The non-dispersive term cannot be expressed in terms of propagating degrees of freedom. Instead, as discussed in Appendix~\ref{misuse}, $\chi_{\rm non-dispersive}$ can be represented in terms of the Veneziano ghost and is saturated by a topologically protected pole at $q^2=0$. This pole explicitly reflects the sensitivity of the non-dispersive contribution to arbitrarily large distances. Physically, this term originates from tunnelling transitions between distinct topological sectors rather than from the propagation of physical states. This also explains why it enters the topological susceptibility with a sign opposite to that of ordinary dispersive contributions. By contrast, $\chi_{\rm dispersive}$ is saturated by physical states separated from the vacuum by a mass gap of order $\Lqcd$ and is therefore insensitive to arbitrarily large distances.

2. The same topological long-range order also manifests itself through corrections to the vacuum energy that are linear in the infrared cutoff. Naively, one might expect finite-size corrections in a gapped theory to be exponentially suppressed. This expectation is indeed valid for contributions associated with propagating states, but not necessarily for the non-dispersive topological component. This can be demonstrated explicitly in exactly solvable models. One example is two-dimensional QED, briefly reviewed in Appendix~\ref{2d_QED}, where the correction to the vacuum energy scales as $\mathbb{L}^{-1}$ despite the presence of a mass gap. Another example is the so-called ``deformed QCD'' model \citep{Thomas:2012ib}, in which the vacuum-energy correction also scales linearly with the inverse size, $\sim\mathbb{L}^{-1}$. This behaviour is analogous to the role played by the parameter $\kappa$ in Eq.~(\ref{vacuum_energy}). A similar effect arises for QCD formulated on hyperbolic space, as discussed below.

3. A central result of \cite{Zhitnitsky:2015dia} is precisely the appearance of a correction linear in $\kappa$, as shown in Eq.~(\ref{vacuum_energy}). This is a crucial ingredient of the present proposal because, from conventional locality arguments, one would instead expect the leading correction at small curvature to be quadratic, $\propto\kappa^2$. Indeed, the only local geometrical difference between $\mathbb{H}^3_{\kappa}\times\mathbb{S}^1_{\kappa^{-1}}$ and $\mathbb{R}^3\times\mathbb{S}^1$---assuming identical circumferences for $\mathbb{S}^1_{\kappa^{-1}}$ and $\mathbb{S}^1$---is the small curvature $\sim\kappa^2$ of the hyperbolic space $\mathbb{H}^3_{\kappa}$, compared with the vanishing curvature of $\mathbb{R}^3$. If the vacuum-energy difference were determined only by local geometrical quantities, one would therefore expect an expansion involving even powers of $\kappa$, i.e., terms proportional to $\kappa^{2n}$. The explicit calculation of \cite{Zhitnitsky:2015dia}, however, produces a term linear in $\kappa$, demonstrating that the relevant contribution cannot be understood solely in terms of local curvature.

4. The origin of this linear correction, $\sim\kappa$, can be traced to vacuum configurations with non-trivial holonomy, which is itself a non-local quantity. The holonomy is defined by

$$
U(\mathbf{x})\equiv{\cal P}\exp\left(i\int_0^{\mathcal T}dx_4\,A_4(x_4,\mathbf{x})\right),
$$

where $\mathcal P$ denotes path ordering. In particular, one considers the Polyakov line evaluated at spatial infinity,
\be
\label{polyakov}
L={\cal P}\exp\left(i\int_0^{\mathcal T}dx_4,
A_4(x_4,|\mathbf{x}|\rightarrow\infty)\right).
\ee
The operator $\Tr L$ characterizes the self-dual configurations that may contribute to the path integral at finite temperature, $T\equiv\mathcal T^{-1}$, including the low-temperature limit $T\rightarrow0$. For the $SU(2)$ gauge group, the holonomy may be parametrized as
\be
\label{holonomy1}
\frac{1}{2}\Tr L=\cos(\pi\nu).
\ee
For integer values of $\nu$, the holonomy belongs to the center of the group, $\frac{1}{2}\Tr L=\pm1$, corresponding to trivial holonomy. The confining value corresponds instead to $\nu=1/2$, for which $\Tr L=0$.

The usual locality argument therefore does not apply to these configurations. As shown explicitly in \cite{Zhitnitsky:2015dia}, the linear correction $\sim\kappa$ is tied to the non-trivial holonomy defined in Eq.~(\ref{holonomy1}), a gauge-invariant quantity that cannot be reduced to the local curvature. In other words, the correction originates from non-local vacuum configurations and cannot be expressed solely in terms of the local curvature $\sim\kappa^2$.

5. At a more technical level, the linear correction $\sim\kappa$ in $\Delta E_{\rm vac}$ can be traced to the different large-distance behaviour of monopole configurations in the two backgrounds. In hyperbolic space $\mathbb{H}^3_{\kappa}$, the monopole fields are exponentially suppressed beyond the characteristic scale $\kappa^{-1}$, whereas no analogous infrared cutoff is present in flat space $\mathbb{R}^3$ \citep{Zhitnitsky:2015dia}. The difference between the two backgrounds therefore arises precisely from the infrared region of the monopole configurations.

6. Another important technical feature that makes this calculation possible is the conformal equivalence between $\mathbb{H}^3_{\kappa}\times\mathbb{S}^1_{\kappa^{-1}}$ and $\mathbb{R}^4$, which ensures that the subtraction procedure entering Eq.~(\ref{vacuum_energy}) is well defined \citep{Zhitnitsky:2015dia}. In particular, the number of zero modes, or equivalently the dimensionality of the moduli space contributing to the path integral, is the same in the two geometries, and the corresponding moduli-space volumes coincide in the small-$\kappa$ limit for a fixed large radius. As a result, the vacuum-energy difference is infrared finite and exhibits the expected extensive scaling with volume, $E\propto\rho V$. This is a non-trivial property of the non-perturbative calculation, since obtaining extensivity generally requires summing over an infinite ensemble of monopole configurations. We refer the reader to \cite{Zhitnitsky:2015dia} for further details.

7. A further technical point concerns the nature of the degrees of freedom involved. As emphasized in Section~\ref{holonomy}, the mechanism responsible for dark energy in this framework is tunnelling between QCD topological sectors, rather than the dynamics of a real propagating scalar field $\phi$. This distinction can be represented formally by introducing an auxiliary field, the so-called Veneziano ghost; further details and references are given in Appendix~\ref{misuse}. Tunnelling contributions enter correlation functions with a sign opposite to that of a conventional propagating scalar field, such as those appearing in quintessence or k-essence models. This unusual sign is the origin of the terminology ``ghost.'' Crucially, however, the Veneziano ghost is an auxiliary, non-propagating field and therefore does not introduce a physical negative-norm degree of freedom or violate the usual principles of quantum field theory. Its role is instead to encode the non-dispersive contact term associated with topological tunnelling, which by construction cannot be represented in terms of physical propagating degrees of freedom.

8. Finally, the non-dispersive contact term is intrinsically infrared in origin and remains sensitive to arbitrarily large distances. In the Veneziano-ghost representation \citep{Veneziano:1979ec}, the long-range topological contribution is encoded by an auxiliary, non-propagating field and manifests itself through a pole at zero momentum, $q^2=0$. This should be contrasted with conventional ultraviolet contact terms, which are associated with arbitrarily short distances, $x^2=0$. The topological contact term is therefore not an ultraviolet artifact and cannot be removed by the usual ultraviolet renormalization procedure; see \cite{Barvinsky:2017lfl} for further discussion.

We conclude by noting that the possibility of long-range effects in gapped gauge theories has been recognized for some time. Lüscher \citep{Luscher:1978rn}, for example, discussed a `secret long-range force'' in quantum field theories with instantons. Its technical origin can already be traced to the instanton analysis of 't~Hooft \citep{tHooft:1976snw}, where an infrared cutoff is required in the treatment of gauge zero modes, and the resulting corrections are power-law rather than exponentially suppressed. The same type of long-range topological order also appears in exactly solvable gapped theories, such as `deformed QCD''~\citep{Thomas:2012ib} and two-dimensional QED. These complementary descriptions reinforce the central point of this Appendix: a mass gap suppresses the long-range propagation of physical states, but does not preclude infrared sensitivity associated with the topological structure of the vacuum. Further details are given in Appendix~\ref{misuse}.

\section{On the interpretation of the so-called Ghost Dark Energy model}
\label{misuse}

The purpose of this Appendix is to clarify the physical interpretation of the vacuum-energy contribution in Eq.~(\ref{eq:P-rho-subtracted}) and to prevent a common misinterpretation of this expression in terms of a conventional dynamical scalar field. As discussed in Section~\ref{holonomy} and Appendix~\ref{sect:technique}, the relevant contribution is non-dispersive and originates from tunnelling between distinct topological sectors of the QCD vacuum. It may be represented by auxiliary topological fields, closely related to the Veneziano ghost \citep{Veneziano:1979ec}, but these fields do not correspond to propagating physical degrees of freedom. This distinction is essential when interpreting the effective equation of state and, in particular, its possible evolution across $\omega_{\rm DE}=-1$.

The connection with the Veneziano ghost motivated the original terminology `Ghost Dark Energy'' (GDE) \citep{Urban:2009vy,Urban:2009yg}. The word `ghost,'' however, should not be understood in the usual sense of a physical propagating degree of freedom with a wrong-sign kinetic term. In exactly solvable models, such as deformed QCD \citep{Zhitnitsky:2013hs}, the corresponding auxiliary topological field can be constructed explicitly and shown to be non-propagating, with no conventional kinetic term. Its role is to encode the non-dispersive topological contribution to the vacuum energy rather than to introduce an additional physical particle.
\footnote{\label{footnote:ghost}
In the Veneziano-ghost description, the topological contribution is associated with a $1/q^2$ pole at zero momentum with a residue of the ``wrong'' sign \citep{Veneziano:1979ec}. In the explicit construction of \cite{Zhitnitsky:2013hs}, this pole does not correspond to a propagating physical state and therefore does not imply the usual ghost-related violations of unitarity or causality.}

A different interpretation has nevertheless been adopted in parts of the phenomenological GDE literature. For example, in \cite{Cai:2010uf} and subsequent studies, the cosmological evolution associated with Eq.~(\ref{eq:P-rho-subtracted}) was analyzed using relations appropriate to a dynamical scalar field $\phi$, such as
\be
\label{c_s}
\omega_{\rm DE}=\frac{\dot{\phi}^2-2V(\phi)}
{\dot{\phi}^2+2V(\phi)},
\qquad
c_s^2\equiv\frac{\dot p_{\rm DE}}{\dot\rho_{\rm DE}}.
\ee
Within a genuine scalar-field theory, quantities of this kind can be used to characterize the dynamics and stability of the propagating degree of freedom. This interpretation, however, does not apply directly to the present framework. The QCD-induced dark-energy contribution is not carried by a propagating scalar field, and the auxiliary topological field used to represent it does not support physical wave-like excitations. Assigning a physical propagation speed to this auxiliary degree of freedom is therefore not meaningful, and stability arguments based on $c_s^2<0$, such as those discussed in \cite{Ebrahimi:2011js,Biswas:2018voo}, cannot be directly applied.

The same distinction is important when $\omega_{\rm DE}<-1$. In a conventional scalar-field framework, such behaviour may be interpreted in terms of a phantom field, and one may attempt to reconstruct a corresponding potential $V(\phi)$, as discussed for example in \cite{Das:2021uom}. This is not the microscopic interpretation of the present model. Here, $\omega_{\rm DE}<-1$ is an effective property of the evolving non-dispersive vacuum contribution and does not arise from the kinetic dynamics of a fundamental phantom field. The crossing of the $\omega_{\rm DE}=-1$ boundary therefore does not, by itself, imply the presence of a propagating ghost or a phantom instability.

The main point of this Appendix is thus that the effective cosmological behaviour associated with Eq.~(\ref{eq:P-rho-subtracted}) should not be interpreted using the standard intuition developed for dynamical scalar-field models. In particular, the occurrence of $\omega_{\rm DE}<-1$ or the crossing of the phantom boundary does not require the introduction of a new propagating degree of freedom. These features instead reflect the time dependence of a non-dispersive vacuum contribution generated by the underlying topological dynamics.

\section{The adiabatic approximation in the two-dimensional QED Schwinger model}
\label{2d_QED}

In this Appendix, we use the two-dimensional QED Schwinger model to investigate how the topological vacuum contribution is modified when the background varies with time. The basic properties of the model were summarized in Section~\ref{sec:lessintuition}. The theory is characterized by a mass gap $m_\gamma$ and by the finite size $\mathbb{L}$ of the system, when topological sectors and instantons can be explicitly constructed, and computations can be carried out. Here, we allow this size to become time dependent, $\mathbb{L}\rightarrow\mathbb{L}(t)$, thereby providing a simple toy model for an expanding Universe. Our goal is to determine how the finite-size contribution to the vacuum energy is modified when the infrared scale of the system changes with time.

The Schwinger model can be solved using several complementary methods. For our purposes, it is particularly useful to follow the path-integral treatment developed in \cite{Sachs:1991en,Sachs:1995dm} and applied in \cite{Urban:2009wb}, in which the theory is formulated on a Euclidean two-dimensional torus characterized by a dimensional scale $\mathbb{L}$ and a dimensionless complex Teichm\"uller parameter $\tau=\tau_1+i\tau_0$. In this formulation, the instanton configurations, together with their zero and non-zero modes, can be constructed explicitly. This makes the model particularly useful as an exactly solvable analogue of the non-perturbative calculations relevant to QCD, including those performed on the hyperbolic background of \cite{Zhitnitsky:2015dia} that lead to Eq.~(\ref{vacuum_energy}).

Both the vacuum energy and the topological susceptibility can be related to the chiral condensate, which in this model takes the form \citep{Urban:2009wb}
\be
\label{chiral_2d}
\langle\bar{\psi}\psi\rangle
\approx
-\frac{m_{\gamma}}{2\pi}e^{\gamma}
\left[
1-\frac{\pi}{m_{\gamma}\mathbb{L}}
\frac{|\tau|-\tau_0}{\tau_0|\tau|}
\right],
\ee
where terms of higher order in $\mathbb{L}^{-1}$ have been neglected. The first term corresponds to the infinite-volume limit, whereas the second represents the leading infrared finite-size correction. Importantly, this correction scales as $\mathbb{L}^{-1}$ despite the presence of the mass gap $m_\gamma$, in agreement with the arguments presented in Section \ref{sec:topology}.

We now allow the size of the system to vary slowly with time and write, over a sufficiently short interval $\Delta t$,
\be
\mathbb{L}(t)=\mathbb{L}
+
\dot{\mathbb{L}} \Delta t.
\ee
For simplicity, we assume
\be
\dot{\mathbb{L}}\ll m_\gamma\mathbb{L},
\ee
so that the characteristic rate of change of the external background is much smaller than the intrinsic dynamical scale $m_\gamma$ of the theory. We also take $\dot{\mathbb{L}}$ to be approximately constant over the interval considered. In this quasi-static regime, for any fixed $\Delta t$ the quantity
\be
\mathbb{L}_0
\equiv
\mathbb{L}+\dot{\mathbb{L}} \Delta t
\ee
can be treated as an approximately constant system size. The leading infinite-volume term in Eq.~(\ref{chiral_2d}) is therefore unchanged, while the infrared correction is modified from $\propto\mathbb{L}^{-1}$ to $\propto\mathbb{L}_0^{-1}$.

To translate this result into the cosmological variables relevant for QCD-induced dark energy, we identify the corresponding physical scales according to
\begin{eqnarray}
\label{correspondence}
m_{\gamma} &\rightarrow& \Lqcd,
\qquad
\dot{\mathbb{L}} \rightarrow
\frac{|\dot{H}|}{\overline{H}^{2}},
\nonumber\\
\Delta t &\rightarrow& H^{-1},
\qquad
\mathbb{L}^{-1} \rightarrow \overline{H}.
\end{eqnarray}
Here, $\overline{H}$ denotes the asymptotic de Sitter value of the Hubble parameter, $H\rightarrow\overline{H}$ as $t\rightarrow\infty$. The factor $\overline{H}^{-2}$ in the identification of $\dot{\mathbb{L}}$ ensures the correct dimensional correspondence. This mapping should be understood as identifying the analogous physical scales in the two systems: $m_\gamma$ with the QCD mass scale, $\mathbb{L}^{-1}$ with the asymptotic infrared scale $\overline{H}$, and the variation of $\mathbb{L}$ with the time dependence of the cosmological expansion.

The finite-size correction relative to the infinite-volume limit is therefore modified according to
\be
\label{correction}
\Delta E_{2d}
&\propto&
\frac{1}
{m_{\gamma}\left(\mathbb{L}
+\dot{\mathbb{L}}\Delta t\right)}
\nonumber\\
&\Rightarrow&
\frac{\rho_{\rm DE}}{\Lambda^4_{\rm QCD}}
\propto
\frac{\overline{H}/\Lqcd}
{1+|\dot{H}|/(H\overline{H})}.
\ee
Equation~(\ref{correction}) makes the origin of the adiabatic condition particularly transparent. When
\be
\frac{|\dot H|}{H}\ll\overline H,
\ee
the denominator approaches unity and the QCD-induced vacuum contribution approaches its pure de Sitter value. This is precisely the adiabatic condition introduced in Eq.~(\ref{eq:adiabatic}).

Conversely, when the adiabatic condition is strongly violated,
\be
\frac{|\dot H|}{H}\gg\overline H,
\ee
the dark-energy contribution becomes
\be
\label{correction1}
\frac{\rho_{\rm DE}}{\Lambda_{\rm QCD}^4}
&\propto&
\left(\frac{\overline{H}}{\Lqcd}\right)
\left(\frac{H\overline{H}}{|\dot{H}|}\right)
\nonumber\\
&=&
\left(\frac{\overline{H}}{\Lqcd}\right)
\left(\frac{\overline{H}}{\omega}\right)
\ll
\left(\frac{\overline{H}}{\Lqcd}\right).
\ee
Here,
\be
\omega\equiv\frac{|\dot H|}{H}\gg\overline H
\ee
is the characteristic frequency associated with the time-dependent perturbation, as introduced following Eq.~(\ref{eq:adiabatic}). The Schwinger-model calculation therefore reproduces the qualitative suppression discussed in Section~\ref{sect:desitter}: when the background evolves on a timescale much shorter than $\overline{H}^{-1}$, the non-dispersive vacuum contribution is suppressed by the additional factor $\overline{H}/\omega\ll1$. Conversely, as the evolution becomes adiabatic, this suppression disappears and the vacuum energy approaches its de Sitter value.

\bibliography{DESI}

\end{document}